\documentclass[reprint,aps,pra,longbibliography]{revtex4-1}
\usepackage[pdftex]{graphicx}
\usepackage{amsmath}
\usepackage{epstopdf}
\usepackage{dcolumn}
\usepackage{natbib}

\begin{document}


\title[Energy-Tunable Quantum Dot with Minimal Fine Structure Created by Using Simultaneous Electric and Magnetic Fields]{Energy-Tunable Quantum Dot with Minimal Fine Structure Created by Using Simultaneous Electric and Magnetic Fields}

\author{M.~A.~Pooley}
 \affiliation{Toshiba Research Europe Limited, Cambridge Research Laboratory, 208~Science~Park, Milton Road, Cambridge, CB4 0GZ, U.K.}
 \affiliation{Cavendish Laboratory, Cambridge University, J.~J.~Thomson Avenue, Cambridge, CB3 0HE, U.K.}
\author{A.~J.~Bennett}
 \email{anthony.bennett@crl.toshiba.co.uk}
 \affiliation{Toshiba Research Europe Limited, Cambridge Research Laboratory, 208~Science~Park, Milton Road, Cambridge, CB4 0GZ, U.K.}
\author{R.~M.~Stevenson}
 \affiliation{Toshiba Research Europe Limited, Cambridge Research Laboratory, 208~Science~Park, Milton Road, Cambridge, CB4 0GZ, U.K.}
\author{I.~Farrer}
 \affiliation{Cavendish Laboratory, Cambridge University, J.~J.~Thomson Avenue, Cambridge, CB3 0HE, U.K.}
\author{D.~A.~Ritchie}
 \affiliation{Cavendish Laboratory, Cambridge University, J.~J.~Thomson Avenue, Cambridge, CB3 0HE, U.K.}
\author{A.~J.~Shields}
 \affiliation{Toshiba Research Europe Limited, Cambridge Research Laboratory, 208~Science~Park, Milton Road, Cambridge, CB4 0GZ, U.K.}

\begin{abstract}
The neutral biexciton cascade of single quantum dots is a promising
source of entangled photon pairs. The character of the entangled
state is determined by the energy difference between the excitonic
eigenstates, known as fine structure splitting (FSS). Here we reduce
the magnitude of the FSS by simultaneously using two independent
tuning mechanisms, in-plane magnetic field and vertical electric
field. We observe that there exists a minimum possible FSS in each
QD which is independent of these tuning mechanisms. However, with
simultaneous application of electric and magnetic fields we show the
FSS can be reduced to its minimum value as the energy of emission is
tuned over several meV with a 5T magnetic field.
\end{abstract}

\maketitle




Semiconductor quantum dots (QDs) confine carriers to nano-sized
regions, resulting in the creation of a discrete set of energy
levels. Optical transitions between these states in InGaAs QDs have
many applications in the fields of quantum optics and optical
quantum computation. In particular, the radiative decay of the
biexciton state ($|X_2\rangle$) via the exciton state
($|X_1\rangle$), has generated significant interest as a source of
on-demand entangled photon pairs\cite{Benson2000,Salter2010}. This
decay process, $|X_2\rangle \rightarrow |X_1\rangle \rightarrow
|0\rangle$, is split into two separate paths by the fine-structure
splitting (FSS, $s$) of the $|X_1\rangle$ state\cite{Bayer2002}. A
finite value of $|s|$ results in the evolution of the $|X_1\rangle$
state during the time between the two emission events, leading to a
time-dependant phase between the two components of the emitted
two-photon state\cite{Stevenson2008}. Therefore, for applications
which require a known input state, such as photonic quantum
computing operations, it is desirable to reduce or eliminate $|s|$.

This has motivated research into methods to manipulate the FSS,
including the application of piezoelectric
strain\cite{Seidl2006,Bryant2010}, intense coherent
lasers\cite{Jundt2008,Muller2009}, magnetic
fields\cite{Bayer1999,Stevenson2006a}, and electric
fields\cite{Vogel2007,Gerardot2007,Kowalik2007,Bennett2010a,Kowalik2006,Marcet2010}.
Alternatively, several groups are pursuing the growth of dots under
particular conditions that naturally give rise to minimal
fine-structure: by targeting particular emission energies on the
(100) surface of GaAs \cite{Salter2010} or the (111)A surface of
GaAs \cite{Kuroda2013}. However, whilst such methods are effective
at reducing $|s|$, several studies report coherent coupling between
the two exciton eigenstates which results in a minimum value, $s_0$
\cite{Singh2010,Bryant2010,Bennett2010a,Trotta2012,Wang2012}.

The use of an electric field orientated parallel to the sample
growth direction is particularly promising, as it allows $|s|$ to be
varied over a range on the order of 100 $\mu$eV and permits
individual QDs to be independently addressed using multiple electric
contacts. In certain samples it is possible to minimise $|s|$ in a
significant proportion of QDs from an ensemble. However, the value
of $s_0$ varies between QDs and is often non-zero.

Recent theoretical and experimental studies have explored the
possibility of manipulating $s_0$ via the simultaneous application
of two independent tuning mechanisms. It has been proposed that
$s_0$ can be reduced to $\sim{0.1}$ $\mu$eV in all InGaAs/GaAs QDs
using two combined strain fields\cite{Wang2012}. Also, it has been
demonstrated that the FSS can be eliminated in such QDs with
simultaneous application of a strain field and an electric
field\cite{Trotta2012}. However, in both cases a single minimum
$s_0$ is obtained at a particular combination of the two tuning
parameters, restricting the emission to a single energy when $s_0 =
0$. In this letter we present a method which allows the emission
energy to be varied whilst maintaining the FSS at its minimum value.
This is achieved via the use of a Voigt magnetic field, orientated
in the plane of the sample, in conjunction with an electric field
which is applied parallel to the sample growth direction. We show
that, by pre-selection of a QD with sufficiently small $s_0$, this
technique may be suitable for the creation of an on-demand
`energy-tunable' source of entangled photons. Such a development is
pre-requisite for building networks of multiple QDs, connected by
photonic interference.

\begin{figure}[htbp]
 \begin{center}
\includegraphics[width = 80mm]{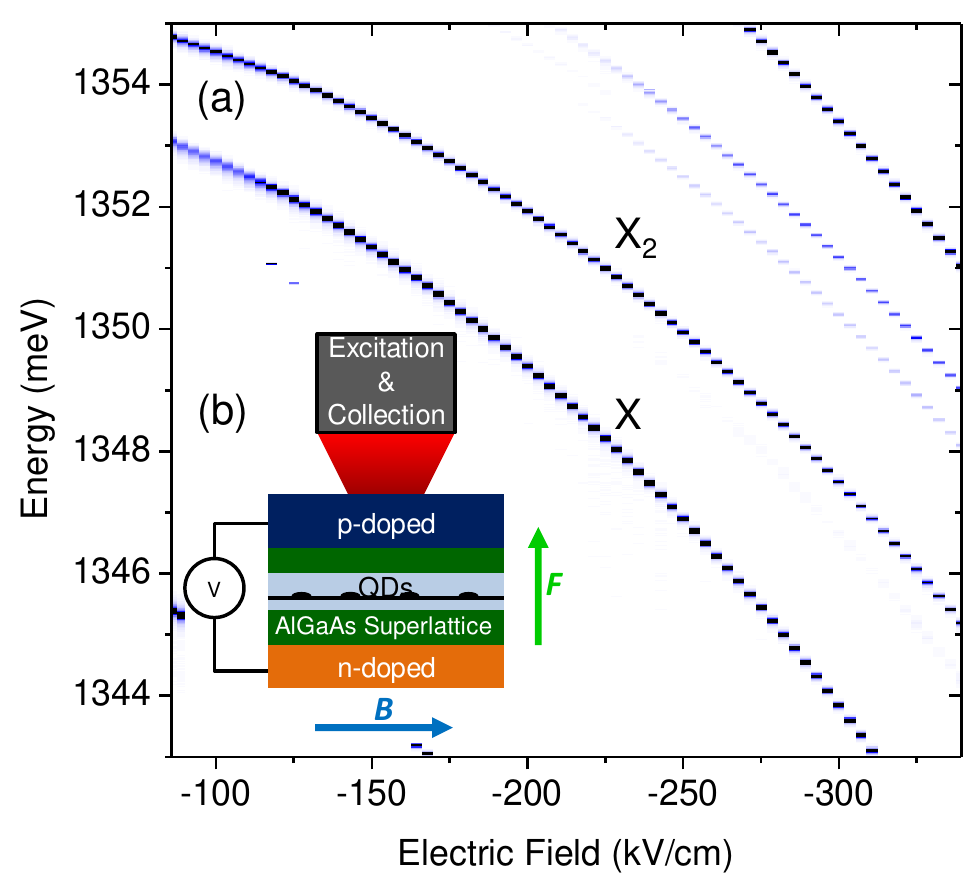}
 \caption{(color online)(a) Photoluminescence emission from the $|X_2\rangle \rightarrow
|X_1\rangle$ and $|X_1\rangle \rightarrow |0\rangle$ transitions as
a function of electric field, $F$, without magnetic field. Spectral
lines are labeled with the initial state of the corresponding
transition. (b) Schematic diagram of device structure showing the
relative orientation of the electric and magnetic fields, $F$ and
$B$ respectively.}
 \label{FIG1}
 \end{center}
\end{figure}

The devices used for this work are p-i-n diodes, as detailed
elsewhere\cite{Bennett2010}, in which the QDs are placed at the
center of the intrinsic region, between two AlGaAs/GaAs superlattice
tunnel barriers, inside a planar cavity with 2 and 13 periods above
and below the QDs, respectively. These devices allow the application
of an electric field in the sample growth direction, leading to
large Stark shifts in the transition energies (Figure
\ref{FIG1}(a)). The transition linewidths remain below the
resolution of our spectrometer across this electric field range
\cite{Bennett2010}. We choose to apply magnetic field perpendicular
to the growth direction, parallel to the [110] crystal axis (Voigt
configuration) using a superconducting magnet, as this has
previously been shown to reduce the FSS in certain dots
\cite{Stevenson2006a}. In contrast, a magnetic field parallel to the
growth direction always leads to an increase in the FSS
\cite{Bayer2002}. Figure \ref{FIG1}(b) shows a schematic diagram of
the device structure along with the orientation of the electric and
magnetic fields. The FSS is extracted from the energy difference
between exciton and biexciton transitions recorded as a function of
polarization angle (as described in \cite{Bennett2010a}). This
technique enables the fine structure to be measured with an error of
$\pm$ 0.5 $\mu$eV, using a spectrometer with a resolution of $\sim$
25 $\mu$eV.  However, at small values of the FSS it is not possible
to separate the polarisation properties of the two neutral
eigenstates but the total emission is isotropic.

\begin{figure*}
 \begin{center}
  \includegraphics{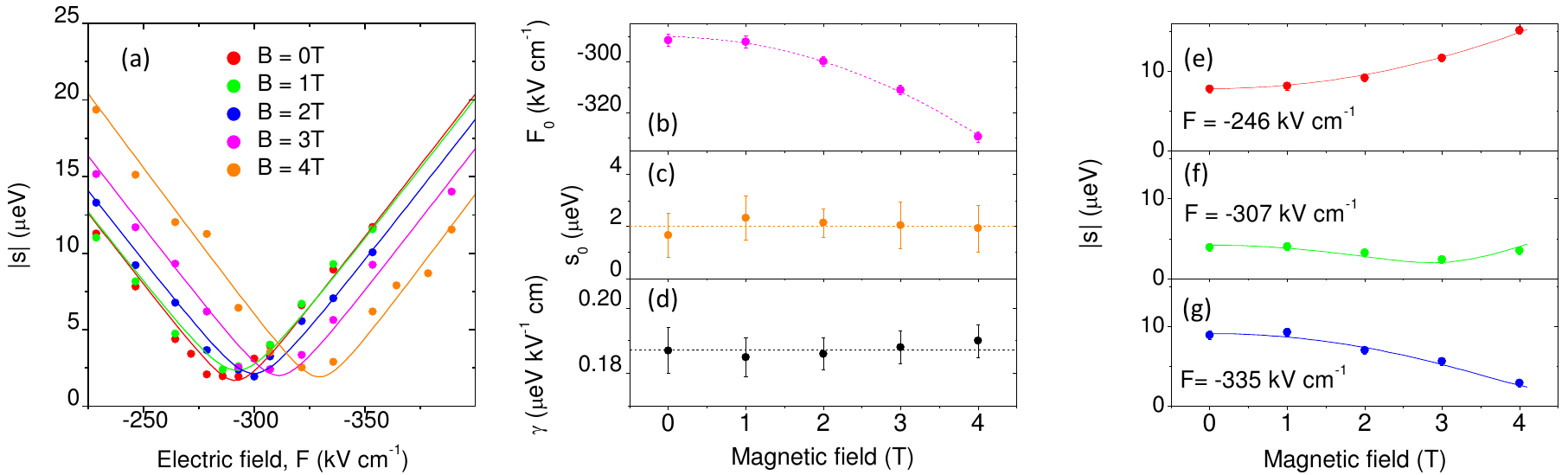}
 \caption{(color online). (a) Magnitude of the fine-structure splitting, $|s|$, as a
function of electric field, $F$, for five different magnetic field
strengths, $B$. Solid lines are fits using equation \ref{EQU:s(E)}.
(b) Electric field at which $|s|=s_0$, $F_0$, as a function of
magnetic field $B$. (c) Minimum $|s|$ as a function of $B$. (d)
Tuning rate of $|s|$ with $F$ as a function of $B$. (e)-(g)
Magnitude of FSS, $|s|$, as a function of $B$ for three different
values of $F$. Solid lines are fits.}
 \label{FIG2}
 \end{center}
\end{figure*}

The behavior of $|s|$ as a function of electric field, $F$, is
described by a hyperbola given by
\begin{equation}\label{EQU:s(E)}
|s|      =      \sqrt{\gamma^2\left(F-F_0\right)^2     +     s_0^2},
\end{equation}
where $\gamma$ is the rate at which $|s|$ varies with $F$ in the
absence of coupling effects, and $F_0$ is the electric field
required to minimise $|s|$\cite{Bennett2010a}. Figure \ref{FIG2}(a)
shows $|s|$ as a function of $F$ for five different magnitudes of
Voigt magnetic field, $B$, for an example QD with an $s_0$ of
$2.0\pm0.2 \, \mu$eV. The solid lines in Figure \ref{FIG2}(a) show
least-squares fits to the data using equation \ref{EQU:s(E)}, from
which the data points and errors for $F_0$, $s_0$ and $\gamma$ are
extracted (Figure \ref{FIG2}(b) - (d)). These figures show that for
a given magnetic field it is possible to minimize the FSS at a
certain electric field $F_0$, but that $\gamma$  and $s_{0}$ are
unchanged by $B$. The constant value of $\gamma$ suggests that a
Voigt magnetic field does not affect the difference in permanent $z$
dipole moment between the two exciton eigenstates\cite{Finley2004}.

This variation of $F_0$ as a function of $B$ can be explained by
considering the additional contribution to the fine structure
splitting due to the Voigt magnetic field, along with the
observation that the coupling strength $s_{0}$ between the
eigenstates is independent of this field. The magnetic field induces
an additional splitting, $\Delta s$, between the two exciton
eigenstates which is well approximated by $\Delta s = \kappa B^2$,
where $\kappa$ is dependent on the in-plane anisotropy of the QD
along with the g-factors of the confined
carriers\cite{Bayer2000,Stevenson2006a, Bennett2013}. Thus, in the
presence of a Voigt magnetic field, the magnitude of $F$ required to
minimize $|s|$ is increased. The value of $F_0$ is then given by
\begin{equation}\label{EQU:F0(B)}
F_0\left(B\right)     =     F_0(0)    -    \frac{\kappa}{\gamma}B^2,
\end{equation}
where $F_0(0)$ is the value of $F_0$ in the absence of the magnetic
field and the second term in this equation is the change in electric
field required to remove the additional, magnetically induced,
component of the fine structure splitting. Figure \ref{FIG2}(b)
shows $F_0(B)$ as a function of magnetic field fit with equation
\ref{EQU:F0(B)} from which a value of $\kappa=0.45\pm0.02 \,
\mu$eV$\,T^{-2}$ is extracted.

The effect of simultaneous application of both the electric and
magnetic field is found by substituting equation \ref{EQU:F0(B)}
into equation \ref{EQU:s(E)}. Figures \ref{FIG2}(e)-(g) show $|s|$
as a function of $B$ for three different values of $F$. There is
good agreement between the experimental measurements and the model.
The value of $|s|$ is either increased or decreased by the
application of $B$, depending on the relative sign of $\kappa$ and
that of the FSS at zero magnetic field, $s(B=0)$. For the QDs
studied here $\kappa>0$, in figure \ref{FIG2}(e) $s(B=0)>0$ leading
to an increase in $|s|$ with $B$; whereas in figure \ref{FIG2}(g)
$s(B=0)<0$, resulting in a reduction of $|s|$. This behavior is
similar to that reported in \cite{Stevenson2006a}. However, Figure
\ref{FIG2}(f) shows data at -301 $kV/cm$ where $B$ causes the fine
structure to pass through a minimum value, $s_0$. This behavior has
not been observed before, and indicates that the avoided crossing in
$s$ is a fundamental property of these QDs and not an artifact of
the tuning mechanism \cite{Bennett2010a, Plumhof2011}.

\begin{figure}[htbp]
 \begin{center}
  \includegraphics[width = 90mm]{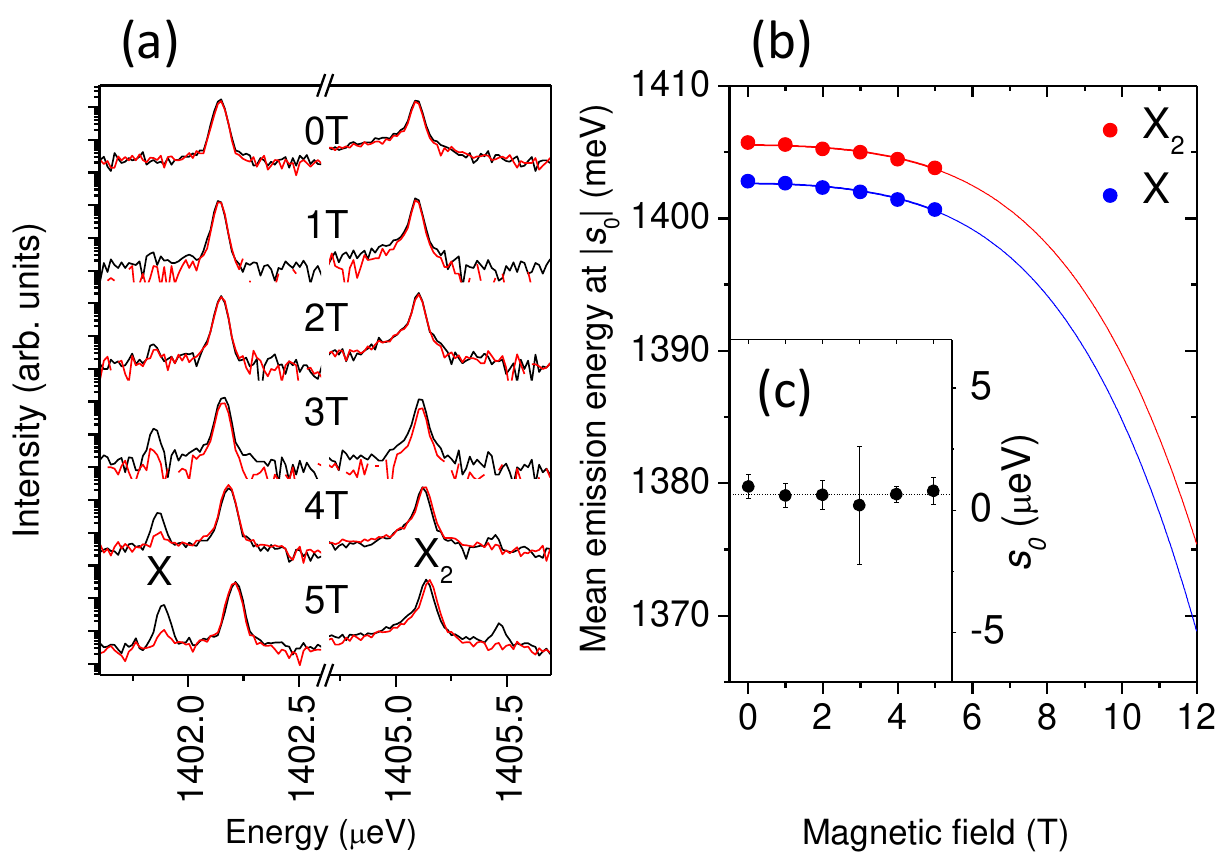}
\caption{(color online). (a) Polarised emission spectra from the
neutral states as a function of magnetic field at $F$ = -100
$kV/cm$. Red and black lines correspond to polarizations along
orthogonal crystal axes. (b) Emission energy
  of the $|X_2\rangle \rightarrow |X_1\rangle$ and $|X_1\rangle \rightarrow
|0\rangle$ transitions when $F=F_0$ as a function of $B$. The solid
lines are fitted and extrapolated to 12T. (c) $s_{0}$ as a function
of $B$ measured up to 5T.}
 \label{FIG3}
 \end{center}
\end{figure}

Figure \ref{FIG3} shows data from the neutral cascade for a second
QD with $s_0=0.6 \pm 0.5 \,\mu$eV, which is below the FSS of those
QDs previously shown to generate entangled photon
pairs\cite{Salter2010,Bennett2010a}. The behavior is similar to that
of all dots we have studied, such as the example in Figure
\ref{FIG2} with a larger $s_{0}$. Polarized spectra from this dot
are shown in Figure \ref{FIG3}(a) at an electric field of $F =-$100
$kV/cm$. The bright excitonic state fine-structure, $s$, passes
through the minimum value at $B \sim$ 2 T as the magnetic field is
changed at this $F$. Negligible change in transition intensities and
widths are observed over this range of magnetic field, within the
resolution of our measurements. At higher magnetic fields the mixing
of dark and bright exciton states \cite{Bayer2000} results in the
appearance of weak emission lines on either side of the bright
exciton states. The 0.4 $meV$ dark-bright splitting we have measured
is comparable to that reported elsewhere for similar dots
\cite{Stevenson2006a} and is sufficiently greater than our spectral
resolution that it does not interfere with measurements of $s$. The
role of the dark-states in driving the change in FSS with magnetic
field has been widely studied \cite{Stevenson2006a,Bayer2000,
Bayer2002}. It has been shown that although the dark-bright mixing
changes the exciton lifetime, entangled photon generation is
preserved because the coherence of the exciton-superposition is much
longer than the radiative lifetimes \cite{Stevenson2008}.

The value of $|s|$ as a function of $F$ for six values of $B$ was
measured and used to determine $s_0$ (Figure \ref{FIG3}c) along with
the emission energy of the two bright neutral transitions at
$F_{0}(B)$, (figure \ref{FIG3}(b)). The energy of the photons
emitted from the bright neutral transitions depends on $F$, due to
the electric field dependent Stark shift described by
\begin{equation}\label{EQU:Stark}
E        =       E_0       -       pF       +       \beta       F^2,
\end{equation}
where $E_0$ is the energy in the absence of an electric field, $p$
is the component of the dipole moment which is parallel with $F$,
and $\beta$ is the polarisability. Therefore, the ability to tune
the value of $F_0$ allows the energy of the photons in the emitted
two-photon state to be varied whilst maintaining the FSS at a the
minimum value of $|s|=s_0$. The energy of the photons emitted at
$F=F_0$ is found by combining equations \ref{EQU:F0(B)} and
\ref{EQU:Stark}. From the parabolic shift of the emission energy,
Stark shift parameters of $p=-6.1 \pm 0.4 \,\mu$eV$\,\text{cm
kV}^{-1}$ and $\beta=0.15 \pm 0.3 \,\mu$eV$\,\text{cm}^2
\text{kV}^{-2}$ are obtained.

The range over which the photon energy can be tuned is dependent on
two factors: the maximum electric field which can be applied without
quenching the optical activity of the QD; and the maximum available
magnetic field which can be applied. For the work presented here,
the tuning range was restricted by the latter. However, the tuning
range is quadratically increased by considering magnetic fields
greater than those available in this study. For the devices studied
in this work, the maximum magnitude of electric field which can be
applied whilst preserving optical emission is $F_{\text{max}}\sim{-
430}$ kV$\,$cm$^{-1}$. At $B=5T$ the electric field required to
minimise $|s|$ is $F_0(B=5T)=-151 \, \text{kV}\,\text{cm}^{-1}$,
which is well below this value. It is possible to increase the
tuning range by using a larger magnetic field until $F_0(B)$ reaches
$F_{\text{max}}$ which occurs at $B=11$ T. Figure \ref{FIG3}(b)
shows the emission energy of the two neutral transitions at $F=F_0$
for $B$ up to 12 T. The maximum tuning range possible with these
devices, which is achieved with a magnetic field of $B=11$ T, is
22.5 meV and 25.4 meV for the $|X_2\rangle \rightarrow |X_1\rangle$
and $|X_1\rangle \rightarrow |0\rangle$ transitions respectively.
This range could be further improved by increasing the amount of
AlGaAs in the device barrier layers, thus reducing carrier tunneling
and increasing $F_{\text{max}}$.

In conclusion, we have presented a method of manipulating $|s|$
using simultaneous application of a Voigt geometry magnetic field
and an orthogonal electric field. In particular, we have
demonstrated that this method may be suitable for the generation of
energy-tunable entangled photon pairs, using QDs selected to have
small $s_0$. Experimentally we achieve a tuning range of a few meV
with a 5T magnetic field but show this could be extended to tens of
meV with an 12T magnet. We expect our technique to find application
to dots emitting at energies where more efficient detectors are
available \cite{Kuroda2013, Marcet2010} and to dots emitting at
wavelengths compatible with the absorption minima in optical fibres
\cite{Ward2014}. A promising future development of this method would
be to incorporate an applied strain field as a third independent
tuning mechanism. As simultaneous application of strain and electric
field has been demonstrated to reduce $s_0$ to zero in any
QD\cite{Trotta2012}, it removes the requirement to pre-select QDs
with small $s_0$.

\bibliographystyle{apsrev4-1}
\bibliography{Pooley_Refs}


\end{document}